\documentclass[journal=jacsat,manuscript=article]{achemso}
\usepackage{chemformula} 
\usepackage[T1]{fontenc} 
\usepackage{epsfig}
\usepackage{graphicx}
\usepackage{titlesec}
\usepackage{dcolumn}
\usepackage{amsthm,amsmath}
\usepackage{comment}
\usepackage[colorlinks]{hyperref}
\usepackage{xcolor}
\usepackage[T1]{fontenc}
\usepackage{mathrsfs}
\usepackage{longtable}
\usepackage{orcidlink}
\usepackage{amsmath}
\usepackage{tabularx}
 \usepackage{tabularray}
\usepackage[normalem]{ulem}

\usepackage{resizegather}
\UseRawInputEncoding

\author{Somesh Chamoli}
\affiliation[Unknown University]
{\small Department of Chemistry, Indian Institute of Technology Bombay, Powai, Mumbai 400076, India}
\author{Xubo Wang}
\affiliation[Unknown University]
{\small Department of Chemistry, the Johns Hopkins University, Baltimore, Maryland 21218, United States}
\author{Chaoqun Zhang}
\affiliation[Unknown University]
{\small Department of Chemistry, Yale University, New Haven, Connecticut 06520, United States}
\author{Malaya K. Nayak}
\affiliation[Unknown University]
{\small Theoretical Chemistry Section, Bhabha Atomic Research Centre, Trombay,
Mumbai 400085, India}
\author{Achintya Kumar Dutta}
\email{achintya@chem.iitb.ac.in}
\affiliation[Unknown University]
{\small Department of Chemistry, Indian Institute of Technology Bombay, Powai, Mumbai 400076, India}

\title[An \textsf{achemso} demo]
  {Frozen natural spinors for Cholesky decomposition based two-component relativistic coupled cluster method}


\begin{document}
\begin{abstract}
  We present an efficient and cost-effective implementation for the exact two-component atomic mean field (X2CAMF)--based coupled cluster (CC) method, which integrates frozen natural spinors (FNS) and the Cholesky decomposition (CD) technique. The use of CD approximation greatly reduces the storage requirement of the calculation without any significant reduction in accuracy. Compared to four-component methods, the FNS and CD-based X2CAMF-CC approach gives similar accuracy as that of the canonical four-component relativistic coupled cluster method at a fraction of the cost. The efficiency of the method is demonstrated by the calculation of a medium-sized uranium complex involving the correlation of over 1000 virtual spinors.
\end{abstract}
\pagebreak
\section{Introduction}
Reliable predictions of molecular properties for molecules containing heavy elements require an accurate treatment of both relativistic effects and electron correlation. The coupled cluster (CC) methods utilizing relativistic four-component or two-component Hamiltonians have proven to be effective approaches for achieving high accuracy in quantum chemistry simulations involving such molecules\cite{PhysRevA.49.1724,https://doi.org/10.1002/qua.560520821,https://doi.org/10.1002/qua.560560844,10.1063/1.472655,LEE199897,10.1063/1.3518712,https://doi.org/10.1002/wcms.1536}. The four-component relativistic coupled cluster (CC) methods, based on the Dirac-Coulomb(-Breit) 
Hamiltonian, are among the most accurate relativistic CC techniques available; however, they are also the most computationally expensive.  The transformation and storage of two-electron integrals in the molecular spinor basis have limited the use of four-component CC methods to relatively small molecular systems.

To alleviate the high computational cost associated with four-component methods, various two-component theories\cite{PhysRevA.33.3742,https://doi.org/10.1002/(SICI)1097-461X(1996)57:3<281::AID-QUA2>3.0.CO;2-U,10.1063/1.473860,NAKAJIMA1999383,BARYSZ2001181,10.1063/1.3159445,saue:hal-00662643} have been introduced. Among these, the exact two-component (X2C) theory\cite{10.1063/1.473860,10.1063/1.3159445,10.1063/1.2137315,10.1063/1.2436882,10.1093/oso/9780195140866.001.0001} has gained recognition as a particularly prominent approach, which can reduce the computational cost of the integral transformation step. The literature covers a diverse array of X2C-CC methods, each providing various approaches and enhancements suited to different applications and computational requirements. This includes the spin-free (SF) version of X2C-CC theory in its mean-field form (the SFX2C-mf scheme)\cite{10.1063/1.5095698}, the SFX2C-CC theory in its one-electron form (the SFX2C-1e scheme)\cite{10.1063/1.3603454,https://doi.org/10.1002/qua.24578,10.1063/1.3624397}, and the X2C molecular mean field (X2CMMF) based CC method\cite{10.1063/1.3239505,10.1063/1.4891801,doi:10.1021/acs.jctc.1c00260,10.1063/1.1413510,doi:10.1021/acs.jpca.3c08167}.
Among various X2C-CC methods, the X2C-CC approach that uses atomic mean field (AMF)\cite{HE1996365} spin-orbit (SO) integrals referred to as the X2CAMF-CC scheme\cite{10.1063/1.5023750,doi:10.1021/acs.jpca.2c02181,10.1063/5.0095112} has emerged as an attractive option to perform highly accurate relativistic coupled cluster calculation. By integrating atomic mean-field spin-orbit effects into the X2C theory, the X2CAMF method delivers a comprehensive treatment of both spin-orbit and scalar-relativistic effects while enhancing computational efficiency and maintaining accuracy for CC calculations, making it particularly suitable for systems involving heavy elements.

Although the X2CAMF-CC scheme provides a computational advantage over the four-component and X2CMMF methods, its application to large or medium-sized molecules remains limited due to the substantial storage required for two-electron integrals.
Furthermore, the inherently high computational cost of the coupled cluster method itself restricts its applicability. For instance, the widely used coupled cluster singles-doubles (CCSD) method scales computationally as O($N^6$) with respect to the size of the basis set. Various approaches, such as density-fitting (DF)\cite{10.1063/1.4807612,10.1063/1.4906344} and Cholesky decomposition (CD)\cite{HELMICHPARIS201938,10.1063/5.0161871,doi:10.1021/acs.jpca.4c04353}, have been employed to reduce the computational cost of relativistic calculations by decomposing the two-electron integrals thereby reducing both disk-space and memory demands. A CD-based X2CAMF-CC and equation of motion (EOM) CC method has also been recently introduced to address storage limitations, enabling calculations for medium-sized molecules\cite{doi:10.1021/acs.jctc.3c01236}.

Despite the advantage of significantly optimized disk and memory requirements, CD-based X2CAMF-CC implementations still involve a similar number of floating-point operations as conventional X2CAMF-CC techniques. Thus, it is crucial to minimize the floating-point operations to make CD-based X2CAMF-CC methods applicable to larger molecules and extend their scope of usefulness. To accelerate floating-point operations, one can utilize massively parallel programs designed to scale across multiple cores\cite{doi:10.1021/ct700268q,doi:10.1021/ct100584w,doi:10.1021/acs.jctc.1c00260}. Alternatively, one can reduce the floating-point operations required in CC calculations by using natural spinors. Natural spinors, the relativistic analog of natural orbitals\cite{PhysRev.97.1474}, offer a more compact description of orbital space and can substantially lower the computational cost of relativistic CD-based X2CAMF-CC calculations. An implementation of frozen natural spinors within both the four-component\cite{10.1063/5.0085932,10.1063/5.0125868,doi:https://doi.org/10.1002/9781394217656.ch5} and two-component\cite{10.1063/5.0087243} frameworks already exists in the literature to reduce the computational cost of relativistic CC and EOM-CC calculations. This paper aims to combine the frozen natural spinor framework and the CD-based X2CAMF-CCSD and CCSD(T) methods to simultaneously reduce storage requirements as well as floating-point operations. The present implementation of the CD-X2CAMF-CC method allows one to routinely perform coupled cluster calculations for systems with more than 1000 spinors in a single computing node.

\section{Theory}
\subsection{Exact two-component Hamiltonian with atomic mean field integrals (the X2CAMF scheme)}
The X2CAMF scheme generates the two-component Hamiltonian by transforming the original four-component Dirac-Coulomb (DC) Hamiltonian using the X2C decoupling approach\cite{10.1063/1.473860}. The four-component DC Hamiltonian can be expressed as 
\begin{equation}
\label{eq:1}
\hat{H^{4c}}=\sum_{pq}{h_{pq}^{4c}}a_{p}^{\dagger}a_{q}+\frac{1}{4}\sum_{pqrs}{g_{pqrs}^{4c}}a_{p}^{\dagger}a_{q}^{\dagger}a_{s}a_{r}
\end{equation}
Here, the summation is restricted to positive-energy spinors only, in accordance with the no-pair approximation\cite{PhysRevA.22.348}, and the indices p, q, r, s denote positive-energy four-component spinors. Within the spin-separation scheme\cite{10.1063/1.466508}, the antisymmetrized two-electron integrals can be split into their spin-free (SF) and spin-dependent (SD) parts
\begin{equation}
\label{eq:2}
g_{pqrs}^{4c}=g_{pqrs}^{4c,SF}+g_{pqrs}^{4c,SD}
\end{equation}
Using Eq. (\ref{eq:2}) in Eq. (\ref{eq:1}) the four-component DC Hamiltonian becomes
\begin{equation}
\label{eq:3}
\hat{H^{4c}}=\sum_{pq}{h_{pq}^{4c}}a_{p}^{\dagger}a_{q}+\frac{1}{4}\sum_{pqrs}{g_{pqrs}^{4c, SD}}a_{p}^{\dagger}a_{q}^{\dagger}a_{s}a_{r}+\frac{1}{4}\sum_{pqrs}{g_{pqrs}^{4c, SF}}a_{p}^{\dagger}a_{q}^{\dagger}a_{s}a_{r}
\end{equation}
Utilizing the local nature of the spin-orbit interaction, the spin-dependent component of the Coulomb interaction in the four-component DC Hamiltonian can be handled using the atomic mean-field (AMF) approximation\cite{HE1996365}.
\begin{equation}
\label{eq:4}
\frac{1}{4}\sum_{pqrs}{g_{pqrs}^{4c, SD}}a_{p}^{\dagger}a_{q}^{\dagger}a_{s}a_{r}\approx\sum_{pq}{g_{pq}^{4c,AMF}}a_{p}^{\dagger}a_{q}=\sum_{pq}\sum_{A}\sum_{i}{n_{i,A}\hspace{0.1cm}g_{pi_{A}qi_{A}}^{4c, SD}a_{p}^{\dagger}a_{q}}
\end{equation}
Here, $A$ represents  the distinct atoms in the molecule, with the index 
$i$ denoting the occupied spinors for atom $A$, and $n_{i,A}$ signifies their corresponding occupation numbers. Therefore, the Hamiltonian becomes 
\begin{equation}
\label{eq:5}
\hat{H^{4c}}=\sum_{pq}{h_{pq}^{4c}}a_{p}^{\dagger}a_{q}+\sum_{pq}{g_{pq}^{4c,AMF}}a_{p}^{\dagger}a_{q}+\frac{1}{4}\sum_{pqrs}{g_{pqrs}^{4c, SF}}a_{p}^{\dagger}a_{q}^{\dagger}a_{s}a_{r}
\end{equation}
Now, the above Hamiltonian can be transformed into a two-component representation. 
In Eq. (\ref{eq:5}), the first term $h_{pq}^{4c}$ has a matrix form of 
\begin{equation}
\label{eq:6}
h^{4c}=\begin{pmatrix}
V & T\\
T & \frac{1}{4m^2c^2}W-T
\end{pmatrix}
\end{equation}
with $V$ as a nuclear potential matrix, $T$ as a kinetic energy matrix, and $W_{pq}=\langle p |(\boldsymbol{\sigma\cdot p})V(\boldsymbol{\sigma\cdot p})| q\rangle$ as small components nuclear attraction matrix, where $\boldsymbol{\sigma}$ are Pauli spin matrices and $\boldsymbol{p}$ is the momentum operator.
In X2C-1e method, as well as X2CAMF scheme, the $h^{4c}$ is transformed into
\begin{equation}
\label{eq:7}
h^{X2C-1e}=R^\dagger[V+X^\dagger T+TX+X^\dagger(\frac{1}{4m^2c^2}W-T)X]R
\end{equation}
where the \textit{X} and \textit{R} are the X2C transformation matrices solved from molecular one-electron Hamiltonian.

In the current implementation of
X2CAMF scheme, the second term in Eq. (\ref{eq:5}) is calculated for each unique atom and transformed into a two-component picture $g_{pq}^{2c,AMF}$ using the atomic $X$ and $R$ matrices. 
The third term in Eq. (\ref{eq:5}), i.e., the scalar two-electron contribution, is approximated by the nonrelativistic two-electron integrals with scalar two-electron picture change error neglected.
\begin{equation}
\label{eq:10}
g_{pqrs}^{4c, SF}\approx g_{pqrs}^{NR}
\end{equation}
Thus, after transforming into a two-component representation, the overall Hamiltonian becomes
\begin{equation}
\label{eq:11}
\hat{H}^{X2CAMF}=\sum_{pq}{h_{pq}^{X2C-1e}}a_{p}^{\dagger}a_{q}+\sum_{pq}{g_{pq}^{2c,AMF}}a_{p}^{\dagger}a_{q}+\frac{1}{4}\sum_{pqrs}{g_{pqrs}^{NR}}a_{p}^{\dagger}a_{q}^{\dagger}a_{s}a_{r}
\end{equation}
The above X2CAMF Hamiltonian can be expressed as an effective one-electron operator combined with a nonrelativistic two-electron operator, as follows:
\begin{equation}
\label{eq:12}
\hat{H}^{X2CAMF}=\sum_{pq}{h_{pq}^{X2CAMF}}a_{p}^{\dagger}a_{q}+\frac{1}{4}\sum_{pqrs}{g_{pqrs}^{NR}}a_{p}^{\dagger}a_{q}^{\dagger}a_{s}a_{r}
\end{equation}
with 
\begin{equation}
\label{eq:13}
h^{X2CAMF}=h^{X2C-1e}+g^{2c,AMF}
\end{equation}
The primary benefit of using the X2CAMF Hamiltonian is that it avoids the formation of molecular relativistic two-electron integrals.

\subsection{Relativistic coupled cluster method}
The foundation of CC theory\cite{Shavitt_Bartlett_2009} lies in the exponential parametrization of the wavefunction
\begin{equation}
\label{eq:14}
|\Psi_{cc}\rangle=e^{\hat{T}}|\Phi_{0}\rangle ,
\end{equation}
where $|\Phi_{0} \rangle$ denotes the reference determinant, and $\hat{T}$ is the cluster excitation operator with its form as
\begin{equation}
\label{eq:15}
\hat{T}=\hat{T_{1}}+\hat{T_{2}}+ \cdots + \hat{T_{N}} ,
\end{equation}
where any general n-tuple excitation operator can be represented as
\begin{equation}
\label{eq:16}
\hat{T_{n}}=\left(\frac{1}{n!}\right)^{2}\sum_{ij...ab...}^{}{t_{ij...}^{ab...}}a_{a}^{\dagger}a_{b}^{\dagger}...\hspace{0.2cm}a_{j}a_{i} \cdots .
\end{equation}
In Eq. (\ref{eq:16}), $t_{ij...}^{ab...}$ denote the cluster amplitudes. The indices $\left(i\hspace{0.1cm},j \hspace{0.1cm},k\hspace{0.1cm}...\right)$ represent occupied spinors, while $\left(a\hspace{0.1cm},b\hspace{0.1cm},c\hspace{0.1cm}...\right)$ indicate virtual spinors. By limiting the cluster operator to include only one-body ($\hat{T_1}$) and two-body ($\hat{T_2}$) excitations, one arrives at the widely used CCSD approximation.
\begin{equation}
\label{eq:17}
\hat{T}=\hat{T_{1}}+\hat{T_{2}}=\sum_{ia}{t_{i}^{a}}a_{a}^{\dagger}a_{i}+\frac{1}{4}\sum_{ijab}{t_{ij}^{ab}}a_{a}^{\dagger}a_{b}^{\dagger}a_{j}a_{i}
\end{equation}
To obtain the cluster amplitudes ($t_{i}^{a}$, and $t_{ij}^{ab}$), one needs to solve a system of non-linear equations simultaneously
\begin{equation}
\label{eq:18}
\langle \Phi_{i}^{a}|\Bar{H}|\Phi_{0}\rangle=0 ,
\end{equation}
\begin{equation}
\label{eq:19}
\langle \Phi_{ij}^{ab}|\Bar{H}|\Phi_{0}\rangle=0 ,
\end{equation}
where $ | \Phi_{i}^{a} \rangle$ and $ | \Phi_{ij}^{ab} \rangle$ are the excited determinants and
$\Bar{H}=e^{-\hat{T}}\hat{H}^{X2CAMF}e^{\hat{T}}$ is the similarity transformed Hamiltonian, with $\hat{H}^{X2CAMF}$ is defined in Eq. (\ref{eq:12}).
The CCSD ground state energy is determined by the expression
\begin{equation}
\label{eq:20}
\langle \Phi_{0}|\Bar{H}|\Phi_{0}\rangle=E, 
\end{equation}
The CCSD(T) method strikes an ideal balance between computational efficiency and accuracy, addressing the limitations of the CCSD method in achieving quantitative precision. In CCSD(T), the triples correction to energy is determined non-iteratively based on the amplitudes from the converged CCSD calculation. 
\begin{equation}
\label{eq:21}
E_{(T)}=\frac{1}{36}\sum_{ijk}\sum_{abc}\left[t(c)_{ijk}^{abc}+t(d)_{ijk}^{abc}\right]D_{ijk}^{abc}\hspace{0.1cm}t(c)_{ijk}^{abc}, 
\end{equation}
with $t(c)_{ijk}^{abc}$ and $t(d)_{ijk}^{abc}$ as connected and disconnected triple amplitudes defined by:
\begin{equation}
\label{eq:22}
D_{ijk}^{abc}\hspace{0.1cm}t(c)_{ijk}^{abc}=P(ij/k)P(ab/c)\left[\sum_{e}t_{ij}^{ae}\langle bc||ek\rangle-\sum_{m}t_{im}^{ab}\langle mc||jk\rangle\right], 
\end{equation}
and
\begin{equation}
\label{eq:23}
D_{ijk}^{abc}\hspace{0.1cm}t(d)_{ijk}^{abc}=P(ij/k)P(ab/c)\left[t_{k}^{c}\langle ab||ij\rangle+t_{ij}^{ab}f_{kc}\right], 
\end{equation}
where
\begin{equation}
\label{eq:24}
D_{ijk}^{abc}=f_{ii}+f_{jj}+f_{kk}-f_{aa}-f_{bb}-f_{cc}, 
\end{equation}
and $P$ is the three-index permutation operator whose action on any arbitrary function $\phi$ is defined as
\begin{equation}
\label{eq:25}
P(pq/r)\phi(pqr)=\phi(pqr)-\phi(rqp)-\phi(prq), 
\end{equation}

\subsection{Cholesky decomposition}
In Cholesky decomposition (CD)\cite{https://doi.org/10.1002/qua.560120408}, any symmetric positive semi-definite matrix ($M$) can be approximated as
\begin{equation}
\label{eq:26}
M\approx LL^{T},
\end{equation}
where $L$ is a lower (or upper) triangular matrix referred to as the Cholesky vectors. This concept can be extended to positive semi-definite electron repulsion integrals (ERIs)
\begin{equation}
\label{eq:27}
(\mu \nu| k\lambda)\approx \sum_{P}^{n_{CD}}L_{\mu \nu}^{P}L_{k\lambda}^{P},
\end{equation}
with $\mu$, $\nu$, $k$, $\lambda$ as AO indices, $L_{\mu \nu}^{P}$ denoting the Cholesky vectors and $n_{CD}$ being the total number of Cholesky vectors. These vectors are generated iteratively by identifying the largest diagonal elements of the ERI matrix $(\mu \nu| \mu \nu)$, and the process continues until the largest diagonal element falls below the predefined Cholesky threshold ($\tau$).

The Cholesky vectors in the AO basis can be transformed to the MO basis as follows.
\begin{equation}
\label{eq:28}
 L_{pq}^{P} = \sum_{\mu \nu}C_{\mu p}^{*}L_{\mu \nu}^{P}C_{\nu q},
\end{equation}
These transformed vectors can then be used to generate antisymmetrized two-electron integrals on the fly in the MO basis.
\begin{equation}
\label{eq:29}
 \langle pq||rs \rangle = \sum_{P}^{n_{CD}}(L_{pr}^{P}L_{qs}^{P}-L_{ps}^{P}L_{qr}^{P}),
\end{equation}
In the current implementation, we avoid constructing and storing integrals of the form $\langle ab||cd \rangle$ and $\langle ab||ci \rangle$, instead generating them on the fly as needed. However, integrals with two or fewer virtual indices are explicitly constructed. 

\subsection{Frozen natural spinors}
Natural spinors are derived from the correlated one-body reduced density matrix (1-RDM), which is constructed from a spin-orbit-coupled wavefunction obtained from a relativistic electron correlation calculation\cite{10.1063/5.0085932}. The eigenfunctions obtained from the diagonalization of this correlated 1-RDM are called natural spinors. They can be identified as the relativistic counterparts of L\"{o}wdin's natural orbitals\cite{PhysRev.97.1474}. The Natural spinors framework is a useful tool for comprehending important spin-dependent phenomena within the confines of the relativistic environment. For instance, natural spinors have the ability to accurately mimic the spin-orbit splitting observed in subshells\cite{10.1063/1.3592780}. In addition, the use of natural spinors can offer insights into the significance of spin-orbit coupling in covalent bonding, as well as aid in the comprehension of the Jahn-Teller effect\cite{doi:10.1021/ct200457q,doi:10.1021/ct300205r}. There is an abundance of research on approaches that utilize natural orbitals for the decrease of computational expenses in non-relativistic correlation calculations\cite{10.1063/1.1696050,10.1063/1.1668917,PhysRevA.1.644,10.1063/1.453884,Taube2005,10.1063/1.3086717,10.1063/1.3173827,10.1063/1.3276630,10.1063/1.3522881,doi:10.1021/acs.jpca.6b11410,10.1063/1.4983277,doi:10.1021/acs.jctc.7b00554,doi:10.1021/acs.jctc.8b00442,10.1063/1.5138643,doi:10.1021/acs.jctc.9b00701,doi:10.1021/acs.jctc.0c01077}. The favorable scaling of MP2 method makes it a popular choice for generating natural orbitals in non-relativistic ground-state coupled cluster calculations. In four-component relativistic methods, the generation of natural spinors for the reduction of computational cost under the no-pair approximation is achieved by utilizing a similar methodology of MP2 theory in the nonrelativistic domain\cite{10.1063/5.0085932,10.1063/5.0087243,10.1063/5.0125868,doi:https://doi.org/10.1002/9781394217656.ch5}.  

To obtain MP2-based natural spinors, one can follow the following steps in sequence: 
First, the virtual-virtual block of the 1-RDM ($D_{ab}$) is constructed using the MP2 method.
\begin{equation}
\label{eq:30}
 D_{ab}=\sum_{cij}^{} {\frac{\langle ac||ij \rangle \hspace{0.1cm}\langle ij||bc \rangle}{\varepsilon_{ij}^{ac} \hspace{0.2cm}\varepsilon_{ij}^{bc}}}
\end{equation}
Here,
\begin{equation}
\label{eq:31}
\varepsilon_{ij}^{ac}=\varepsilon_{i}+\varepsilon_{j}-\varepsilon_{a}-\varepsilon_{c}
\end{equation}
\begin{equation}
\label{eq:32}
\varepsilon_{ij}^{bc}=\varepsilon_{i}+\varepsilon_{j}-\varepsilon_{b}-\varepsilon_{c}
\end{equation}
In Eq. (\ref{eq:31}) and (\ref{eq:32}), $\varepsilon_{i}$, $\varepsilon_{j}$, $\varepsilon_{a}$, $\varepsilon_{b}$, and $\varepsilon_{c}$ are molecular spinor energies, and $\langle ac||ij \rangle$ and $\langle ij||bc \rangle$ denotes the antisymmetrized two-electron integrals. The next step is the diagonalization of $D_{ab}$.
\begin{equation}
\label{eq:33}
D_{ab}V=Vn
\end{equation}
The eigenvectors ($V$) obtained in Eq. (\ref{eq:33})  are called virtual natural spinors, and the corresponding eigenvalues ($n$) are known as occupation numbers. The virtual natural spinors can be categorized based on their relevance to the overall correlation energy by utilizing their respective occupation numbers. By setting a predetermined threshold or cutoff for the occupation numbers of virtual natural spinors, the virtual space can be shortened by retaining only those spinors that meet or exceed this threshold. The subsequent step involves converting the virtual-virtual block of the Fock matrix into the truncated basis of natural spinors.
\begin{equation}
\label{eq:34}
F_{vv}^{NS}=\tilde{V}^{\dagger}F_{vv}\tilde{V}
\end{equation}
Here, $\tilde{V}$ are virtual natural spinors in a truncated basis. $F_{vv}$ represents a virtual-virtual block of the initial canonical Fock matrix, and $F_{vv}^{NS}$ is the virtual-virtual block of the Fock matrix in a truncated virtual natural spinor basis. Afterward, the $F_{vv}^{NS}$ block is diagonalized in the natural spinor basis to obtain the semi-canonical virtual natural spinors  ($\tilde{Z}$ ) and their associated orbital energies ($\epsilon$).
\begin{equation}
\label{eq:35}
F_{vv}^{NS}\tilde{Z}=\tilde{Z}\epsilon
\end{equation}
A transformation matrix ($U$) can be constructed with the help of $\tilde{V}$ and $\tilde{Z}$, which can be used to convert the canonical virtual spinor space to the semi-canonical natural virtual spinor space.
\begin{equation}
\label{eq:36}
U=\tilde{Z}\tilde{V}
\end{equation}
The occupied spinors within our chosen basis are kept in their canonical form, while the virtual spinors undergo a transformation to become semi-canonical natural virtual spinors. Hence, the process is also known as the ``frozen natural spinors (FNS) method''. Using this FNS basis, it is now feasible to conduct higher-order correlation calculations (such as CCSD/CCSD(T)) at low computational costs. 
Coupled cluster methods that utilize natural orbitals in the non-relativistic realm have seen a significant improvement in accuracy by means of perturbative corrections\cite{Taube2005,10.1063/1.3173827}. Similar is the case found in the four-component relativistic coupled cluster method based on frozen natural spinors\cite{10.1063/5.0085932}. By utilizing the $\Delta E_{MP2}$ approximation for $\Delta E_{CCSD/CCSD(T)}$, it is possible to apply a perturbative correction for the truncated virtual space. 
\begin{equation}
\label{eq:37}
\Delta E_{CCSD/CCSD(T)}=E^{Canonical}_{CCSD/CCSD(T)} - E^{FNS}_{CCSD/CCSD(T)}
\end{equation}
\begin{equation}
\label{eq:38}
\Delta E_{CCSD/CCSD(T)}\approx\Delta E_{MP2}
\end{equation}
with
\begin{equation}
\label{eq:39}
\Delta E_{MP2}=E^{Canonical}_{MP2}-E^{FNS}_{MP2}
\end{equation}
\section{Implementation and Computational Details}
The FNS-CD-X2CAMF-CCSD/CCSD(T) method has been implemented in the development version of BAGH\cite{dutta2023bagh}.  The X2CAMF-HF calculations are performed using the \texttt{socutils} package\cite{socutils} interfaced with BAGH. Additionally, only the LOO and LOV-type 3 center two-electron integrals are created in the canonical molecular spinor basis; LVV-type Cholesky vectors are formed directly in the FNS basis, where O and V denote occupied and virtual spinors, respectively. It significantly accelerates the integral transformation phase and dramatically minimizes the storage requirement. Figure 1 presents a schematic depiction of the steps involved in the FNS-CD-X2CAMF-CCSD/CCSD(T) method

\begin{figure}[ht]
    \begin{center}
    \includegraphics[width=0.9\textwidth]{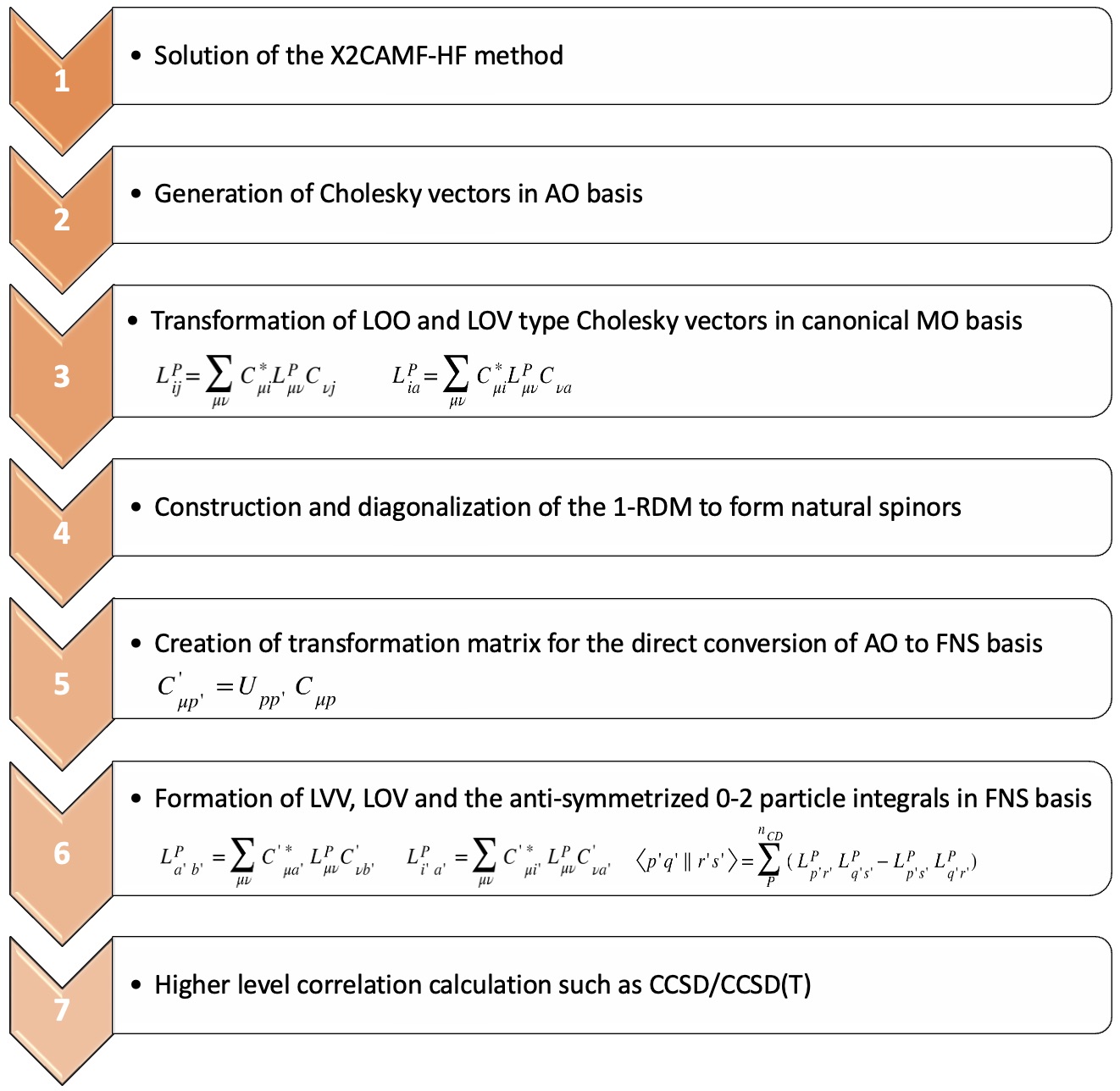}
    \caption{A schematic description of the FNS-CD-X2CAMF-CCSD/CCSD(T) algorithm.}
    \label{fig:my_label1}
    \end{center}
\end{figure}

The first step comprises the solution of the X2CAMF-HF method. Next, the Cholesky vectors are produced in the AO basis.
The next step is to generate LOO and LOV-type Cholesky vectors in a canonical MO basis (using Eq. (\ref{eq:28})), which is necessary for the MP2 calculations. This step is followed by the construction of 1-RDM using MP2 amplitudes and diagonalization of 1-RDM to form natural spinors.
After that, a transformation matrix is generated using Eq. (\ref{eq:36}) and its contraction with the original coefficient matrix to transform the AO basis directly to the FNS basis. The next step is the formation of LVV and LOV-type Cholesky vectors and antisymmetrized 0-2 particle (external) integrals using them in the FNS basis through the transformation matrix. Any construction or storage of four-particle or three-particle one-hole integrals is completely avoided and is generated on the fly using Cholesky vectors in the FNS basis. Now, these transformed integrals can be used for any higher-level correlation calculations, such as CCSD and CCSD(T).

\begin{figure}[H]
    \begin{center}
    \includegraphics[width=1.0\textwidth]{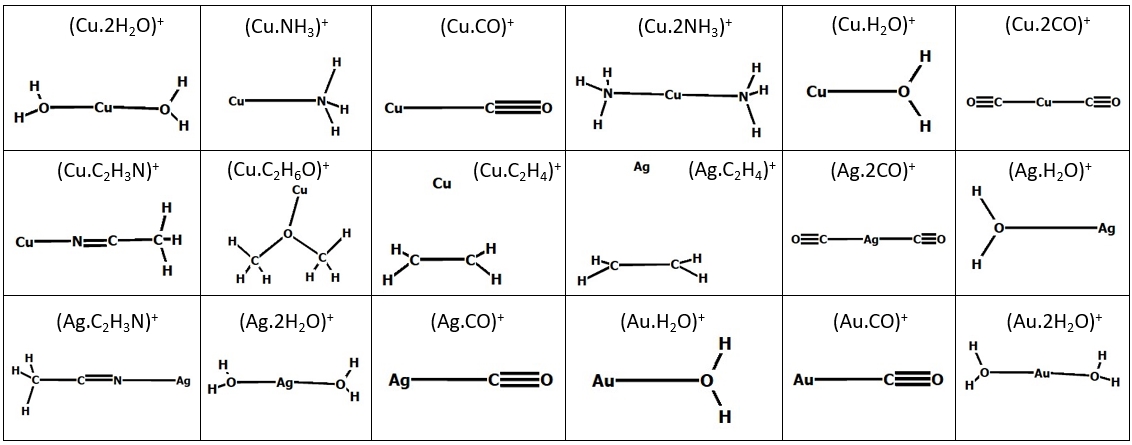}
    \caption{Molecular structures of the complexes used in the current work.}
    \label{fig:my_label2}
    \end{center}
\end{figure}

To evaluate the performance of the FNS-CD-X2CAMF-CCSD/CCSD(T) method, a comprehensive analysis of the noncovalent dissociation enthalpies of ligands in 18 different complexes of coinage metal cations (Cu$^{+}$, Ag$^{+}$, and Au$^{+}$) in the gas phase for which credible experimental studies are available were carried out (See Figure \ref{fig:my_label2} ). Optimized geometries for all the systems are taken from the work of Cavallo and coworkers\cite{doi:10.1021/acs.jctc.5b00584}. For the ligands, the basis set employed was aug-cc-pVXZ (where X = D, T, and Q); on the other hand, for the metal cations (Cu$^{+}$, Ag$^{+}$, and Au$^{+}$), the basis set used was dyall.aexz (where x=2, 3, and 4). The basis sets were kept uncontracted, and the frozen core approximation was utilized for all the calculations. The gas phase noncovalent dissociation enthalpy ($\Delta H^{0}$) of a ligand ($L$) from a metal-ligand complex ($ML_{2}$) for a particular dissociation of type $ML_{2}\rightarrow ML + L$ is calculated as:  
\begin{equation}
\label{eq:40}
\Delta H^{0}=[E_{0}+H_{corr}]_{ML}+[E_{0}+H_{corr}]_{L}-[E_{0}+H_{corr}]_{ML_{2}}
\end{equation}
Where $E_{0}$ is the total electronic energy obtained after the solution of the FNS-CD-X2CAMF-CCSD/CCSD(T) method, and $H_{corr}$ is the enthalpic correction added to the total electronic energy to determine the dissociation enthalpy. Our study will rely on the $H_{corr}$ values provided in the work of Cavallo and coworkers\cite{doi:10.1021/acs.jctc.5b00584} for different systems.
By utilizing the basis extrapolation method, both the basis set superposition error (BSSE) and basis set incompleteness error (BSIE) can be evaded as with the extrapolation scheme; we move closer to the point of achieving a complete basis set (CBS) limit. In our current study, we employed the three-point extrapolation scheme proposed by Peterson and Dunning\cite{10.1063/1.466884} to obtain the reference and correlation energy at the CBS limit. To calculate the dissociation enthalpy of a ligand from a metal-ligand complex at the CBS limit, we use Eq. (\ref{eq:40}) by plugging in the CBS limit values for DHF and correlation energies while keeping $H_{corr}$ consistent with the previously reported value\cite{doi:10.1021/acs.jctc.5b00584}.
\section{Results and discussion}

\subsection{Benchmark calculations for non-covalent bond dissociation enthalpy}
Choosing suitable criteria to truncate the natural spinor virtual space holds the utmost significance. To ensure consistency across different atoms and molecules, the truncation framework should be uniformly applicable while also being seamlessly accessible as a black box. It has been well established in the literature the occupation of natural orbitals in the non-relativistic domain\cite{10.1063/1.3276630,10.1063/1.3173827} and natural spinors in the relativistic domain\cite{10.1063/5.0085932,10.1063/5.0087243,10.1063/5.0125868,doi:https://doi.org/10.1002/9781394217656.ch5} is a credible criterion for arranging the natural orbitals/spinors in order of importance. Hence, in this particular study, we resort to the occupation number of natural spinors as a measure of criteria to truncate the virtual natural spinor space. By adjusting the predefined Cholesky threshold, one can optimize both storage demands and accuracy. To find the optimal threshold that balances computational efficiency with accuracy, we considered three truncation thresholds:  LOOSEFNS (FNS threshold: $10^{-4}$, and CD threshold: $10^{-3}$), NORMALFNS(FNS threshold: $10^{-4.5}$, and CD threshold: $10^{-4}$), and TIGHTFNS(FNS threshold: $10^{-5}$, and CD threshold: $10^{-5}$).
We conducted CCSD and CCSD(T) calculations on 18 different metal-ligand complexes to determine which threshold is most appropriate for calculating non-covalent dissociation enthalpies. For the calculation, we chose the uncontracted aug-cc-pVDZ basis set for the ligands and the dyall.ae2z basis set for the metal. For each threshold, mean error (ME), mean absolute error (MAE), absolute maximum error (AME), and standard error (SE) were calculated. Cheng and co-workers\cite{doi:10.1021/acs.jctc.3c01236} have shown that a Cholesky threshold of $10^{-5}$ gives very close results as that obtained using standard integrals version. We have also investigated the error introduced due to the Cholesky decomposition of integrals (See supporting information) and found the threshold of $10^{-5}$ gives nearly identical results as that of the X2CAMF-CC method with standard integrals. Therefore, the canonical CD-X2CAMF-CCSD/CCSD(T) results with full virtual space and Cholesky threshold of $10^{-5}$ were chosen as the reference. Table \ref{table:1} shows the comparison of errors for CCSD and CCSD(T) methods for 18 metal-ligand complexes in three different thresholds. The error values are shown in Table \ref{table:1}. It can be seen that on moving from a LOOSEFNS to a NORMALFNS threshold, the error is significantly reduced in both the CCSD and CCSD(T) methods. The AME is reduced from 0.890 kcal/mol to 0.340 kcal/mol when going from LOOSEFNS to NORMALFNS. For the CCSD(T) method, the decrease in AME is 0.570 kcal/mol to 0.287 kcal/mol. The change from NORMALFNS to TIGHTFNS is comparatively smaller. The MAE observed in the FNS-CD-X2CAMF-CCSD(T) method with a TIGHTFNS threshold is 0.170 kcal/mol. The NORMALFNS makes a good compromise between computational cost and accuracy. Therefore, the larger basis set calculations are performed using the NORMALFNS threshold only.  

\begin{figure}[H]
    \begin{center}
    \includegraphics[width=1.0\textwidth]{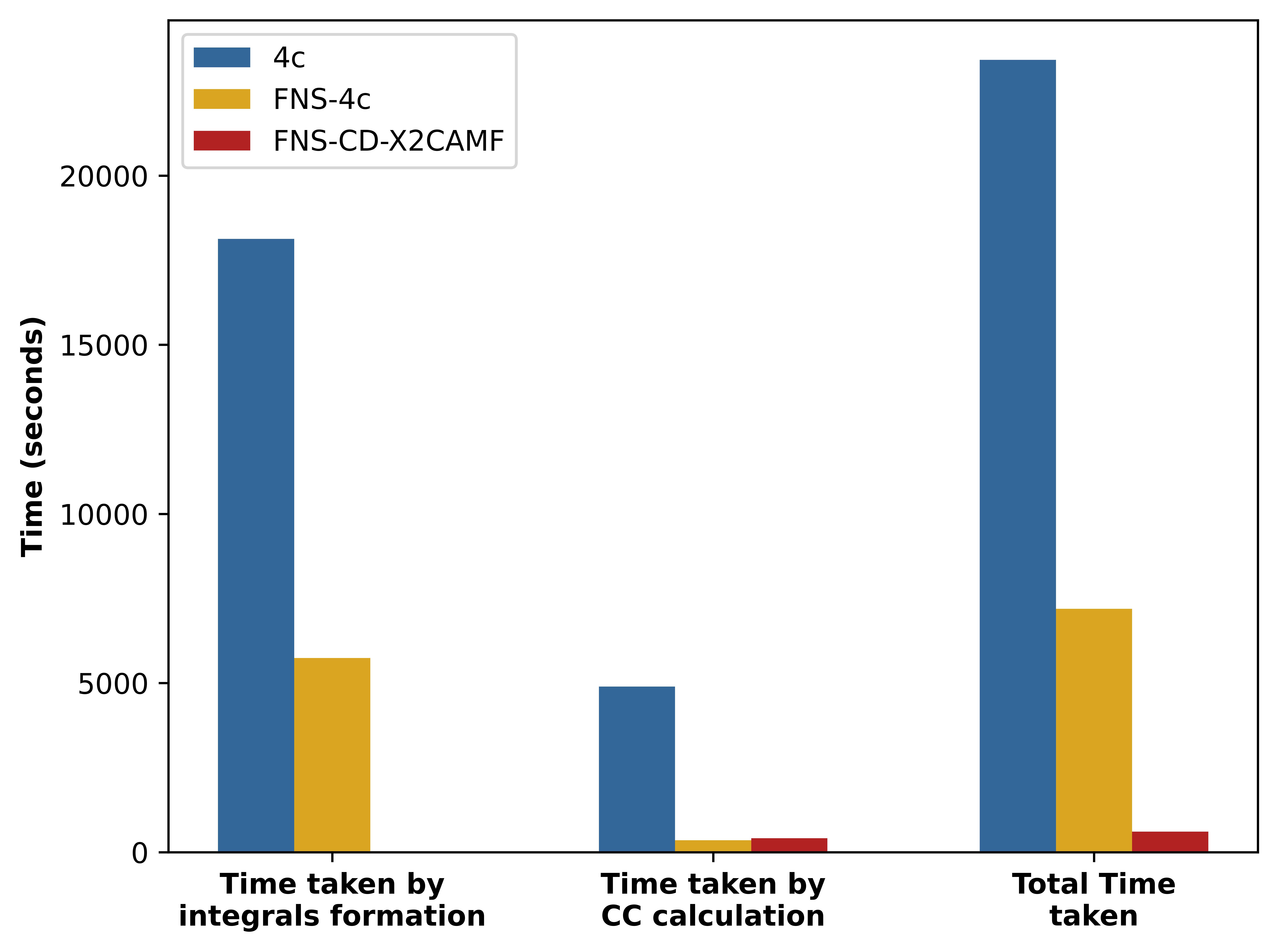}
    \caption{Experimental (with the error bars) and FNS-CD-X2CAMF-CCSD(T) bond dissociation enthalpies of ligand obtained for 18 different metal-ligand complexes at different basis sets.}
    \label{fig:my_label3}
    \end{center}
\end{figure}

We have calculated the non-covalent bond dissociation enthalpies of ligands for the same 18 metal-ligand complexes in three different basis sets using the FNS-CD-X2CAMF-CCSD(T) method. The calculated energy values at aug-cc-pVDZ (dyall.ae2z for the metal), aug-cc-pVTZ (dyall.ae3z for the metal), and aug-cc-pVQZ (dyall.ae4z for the metal) level are extrapolated to obtain dissociation enthalpy at the CBS limit. The absolute values of the dissociation enthalpies in different bases are provided in Table S1. Figure \ref{fig:my_label3} shows the enthalpy values in different basis sets calculated using the FNS-CD-X2CAMF-CCSD(T) method. The figure also includes the corresponding experimental values\cite{doi:10.1021/ja00119a023,doi:10.1021/ja00087a044,doi:10.1021/ja973202c,doi:10.1021/ja973834z,doi:10.1021/jp0132432,doi:10.1021/jp003509p,doi:10.1021/jp014005+,GUO199116,10.1063/1.443497,doi:10.1021/jp002676m,doi:10.1021/jp025557a,https://doi.org/10.1002/anie.200300572} (with associated error bars) for comparison.  From figure \ref{fig:my_label3}, it can be seen that dissociation enthalpy values at QZ and CBS limit are in good agreement with each other and less than 0.5 kcal/mol. The FNS-CD-X2CAMF-CCSD(T) results for a few complexes are a little off the experimental error bar. It is presumably due to the missing effect from higher-order excitation in the coupled cluster. Overall, we find the accuracy of the FNS-CD-X2CAMF-CCSD(T) method using normal threshold and at CBS limit to be reasonable in reproducing experimentally determined gas phase non-covalent dissociation enthalpies of ligands from coinage metal cation complexes.

\begin{table}[H]
\centering
\caption{Comparison of Mean Error (ME), Mean Absolute Error (MAE), Standard Error (SE), and Absolute Maximum Error (AME) (in kcal/mol) of FNS-CD-X2CAMF-CCSD and FNS-CD-X2CAMF-CCSD(T) methods with respect to their canonical values in the different truncation thresholds. The uncontracted aug-cc-pVDZ basis set is used for the ligands and the dyall.ae2z basis set is used for the metal.}
\begin{tabular}{cccccccccc}
\hline \hline \\
  & \multicolumn{4}{c}{FNS-CD-X2CAMF-CCSD} & & \multicolumn{4}{c}{FNS-CD-X2CAMF-CCSD(T)} \\
\cline{2-5} \cline{7-10} \\
 Threshold & ME & MAE & SE & AME & & ME & MAE & SE & AME \\
 \hline \\
 LOOSEFNS  & 0.173 & 0.354 & 0.398 & 0.890 & & 0.146 & 0.298 & 0.301 & 0.570 \\
 NORMALFNS & 0.018 & 0.135 & 0.170 & 0.340 & & 0.093 & 0.128 & 0.117 & 0.289 \\
 TIGHTFNS  & 0.021 & 0.107 & 0.135 & 0.250 & & 0.000 & 0.071 & 0.091 & 0.170 \\
\hline \hline
\end{tabular}
\label{table:1}
\end{table}

\subsection{Comparision with four-component FNS framework }
 To demonstrate the accuracy of the FNS-CD-X2CAMF method with respect to the four-component variant, we calculated the spectroscopic constants (bond length and harmonic vibrational frequency) for the hydrogen halide series(HX, X=F, Cl, Br, and I) using the FNS-CD-X2CAMF-CCSD and FNS-CD-X2CAMF-CCSD(T) methods. This energy derivative has been calculated through numerical differentiation of the total energy, utilizing the TWOFIT utility program in DIRAC software\cite{DIRAC22}. A fifth-degree polynomial was employed, and the basis set used was dyall.acv3z for the Br and I atoms and uncontracted aug-cc-pVTZ for the H, F, and Cl atoms. The same example was used to determine the accuracy of our original four-component FNS implementation\cite{10.1063/5.0085932}.
Tables \ref{table:2} and \ref{table:3} display the errors in bond length and harmonic vibrational frequency for hydrogen halides relative to the canonical four-component values across three truncation thresholds. The tables also provide a comparison with the error values from the four-component FNS method. The canonical four-component values and FNS four-component error values are taken from our previous work\cite{10.1063/5.0085932}. All the error values presented in Tables \ref{table:2} and \ref{table:3} are perturbatively corrected for the FNS truncation. The FNS truncation used in ref \cite{10.1063/5.0085932} is $10^{-5}$, which is the same as the TIGHTFNS setting in the present manuscript. Table \ref{table:2} clearly shows that, with a tight threshold, the error values in the FNS-CD-X2CAMF-CCSD method are of a similar order of magnitude to those in the FNS-4c-CCSD method. This pattern is also observed in the FNS-CD-X2CAMF-CCSD(T) method. 
In our four-component-based implementation, we found that a threshold of 10$^{-5}$ was adequate for obtaining converged results for the bond lengths of the hydrogen halide series. However, in the current implementation, we recommend using the NORMALFNS threshold as a truncation threshold, as it provides error values of a similar order of magnitude to those of a TIGHTFNS threshold while balancing accuracy and efficiency.
Just like the bond lengths, the order of magnitude of the harmonic vibrational frequency error values at the TIGHTFNS threshold for the FNS-CD-X2CAMF-CCSD and FNS-CD-X2CAMF-CCSD(T) methods is comparable to the error values from the FNS-4c-CCSD and FNS-4c-CCSD(T) methods, except the case in HI where the FNS-CD-X2CAMF variant works little better. The harmonic frequency results for all molecules in the FNS-4c-CCSD and FNS-4c-CCSD(T) methods are within 10 cm$^{-1}$ of their canonical counterparts. While the error in bond lengths for the LOOSEFNS threshold was not significant, it is more pronounced for the harmonic frequency values. The maximum deviation of the harmonic frequency values from the canonical 4c method exceeds 11 cm$^{-1}$. The NORMALFNS threshold gives an economic compromise with error  within 5 cm$^{-1}$ of the canonical 4c values,
\begin{table}[H]
\normalsize
\centering
\caption{Comparison of error in bond length (in \AA), with respect to the canonical four-component values (4c) in the different truncation thresholds. The basis set used was dyall.acv3z for the Br and I atoms and uncontracted aug-cc-pVTZ for the H, F, and Cl atoms.}
\resizebox{\textwidth}{!}{  
\begin{tblr}{colspec={p{1.5cm} p{1.5cm} p{1.5cm} p{1.5cm} p{1.5cm} p{1.5cm} p{0.2cm} p{1.5cm} p{1.5cm} p{1.5cm} p{1.5cm} p{1.5cm}},  
  cell{1}{2} = {c=5}{c},
  cell{1}{8} = {c=5}{c},
  cell{2}{4} = {c=3}{c}, 
  cell{2}{10} = {c=3}{c}, 
  cell{3}{4} = {c=1}{c}, 
  cell{3}{10} = {c=1}{c}, 
  hline{2} = {2-6,8-12}{},
  hline{3} = {4-6,10-12}{}
}
\hline \hline 
         & CCSD &        &               &        &       &  & CCSD(T) &        &               &        &       \\
Molecule & 4c   & FNS-4c & FNS-CD-X2CAMF &        &       &  & 4c      & FNS-4c & FNS-CD-X2CAMF &        &       \\
         &      &        & LOOSE         & NORMAL & TIGHT &  &         &        & LOOSE          & NORMAL & TIGHT \\
\hline
HF       & 0.9183 & 0.0000 & 0.0008 & 0.0003 & 0.0001  &  & 0.9213 & 0.0000 & 0.0002 & 0.0002 & -0.0001 \\
HCl      & 1.2753 & 0.0000 & 0.0002 & 0.0000 & -0.0001 &  & 1.2777 & 0.0000 & 0.0002 & -0.0001 & -0.0001 \\
HBr      & 1.4089 & 0.0005 & 0.0004 & 0.0004 & 0.0001  &  & 1.4117 & 0.0003 & 0.0002 & 0.0002  & 0.0000 \\
HI       & 1.6027 & 0.0004 & 0.0009 & 0.0005 & 0.0003  &  & 1.6059 & 0.0004 & 0.0011 & 0.0004  & 0.0002 \\
\hline \hline 
\end{tblr}
}
\label{table:2}
\end{table}

\begin{table}[H]
\normalsize
\centering
\caption{Comparison of error in harmonic vibrational frequency (in cm$^{-1}$),  with respect to the canonical four-component values (4c) in the different truncation thresholds. The basis set used was dyall.acv3z for the Br and I atoms and uncontracted aug-cc-pVTZ for the H, F, and Cl atoms.}
\resizebox{\textwidth}{!}{  
\begin{tblr}{colspec={p{1.5cm} p{1.5cm} p{1.5cm} p{1.5cm} p{1.5cm} p{1.5cm} p{0.2cm} p{1.5cm} p{1.5cm} p{1.5cm} p{1.5cm} p{1.5cm}},  
  cell{1}{2} = {c=5}{c},
  cell{1}{8} = {c=5}{c},
  cell{2}{4} = {c=3}{c}, 
  cell{2}{10} = {c=3}{c}, 
  cell{3}{4} = {c=1}{c}, 
  cell{3}{10} = {c=1}{c}, 
  hline{2} = {2-6,8-12}{},
  hline{3} = {4-6,10-12}{}
}
\hline \hline 
         & CCSD &        &               &        &       &  & CCSD(T) &        &               &        &       \\
Molecule & 4c   & FNS-4c & FNS-CD-X2CAMF &        &       &  & 4c      & FNS-4c & FNS-CD-X2CAMF &        &       \\
         &      &        & LOOSE         & NORMAL & TIGHT &  &         &        & LOOSE         & NORMAL & TIGHT \\
\hline
HF       & 4083.08 & -0.08 & -11.7 & -4.54 & 0.07  &  & 4036.44 & 1.06 & -3.6 & -3.22 & 1.18 \\
HCl      & 3015.13 & -0.03 & -3.06 & 0.41 & -0.02 &  & 2990.66 & 0.24 & -3.25 & 1.23 & 0.21 \\
HBr      & 2673.1 & -0.63 & -0.64 & -3.08 & -0.97  &  & 2647.42 & -0.03 & 1.24 & -0.82  & -0.4 \\
HI       & 2333.7 & -9.09 & -10.32 & -3.39 & -2.32  &  & 2306.75 & -7.59 & -8.41 & -1.96  & -2.71 \\
\hline \hline 
\end{tblr}
}
\label{table:3}
\end{table}

\subsection{Computational efficiency}
\begin{figure}[H]
    \begin{center}
    \includegraphics[width=0.8\textwidth]{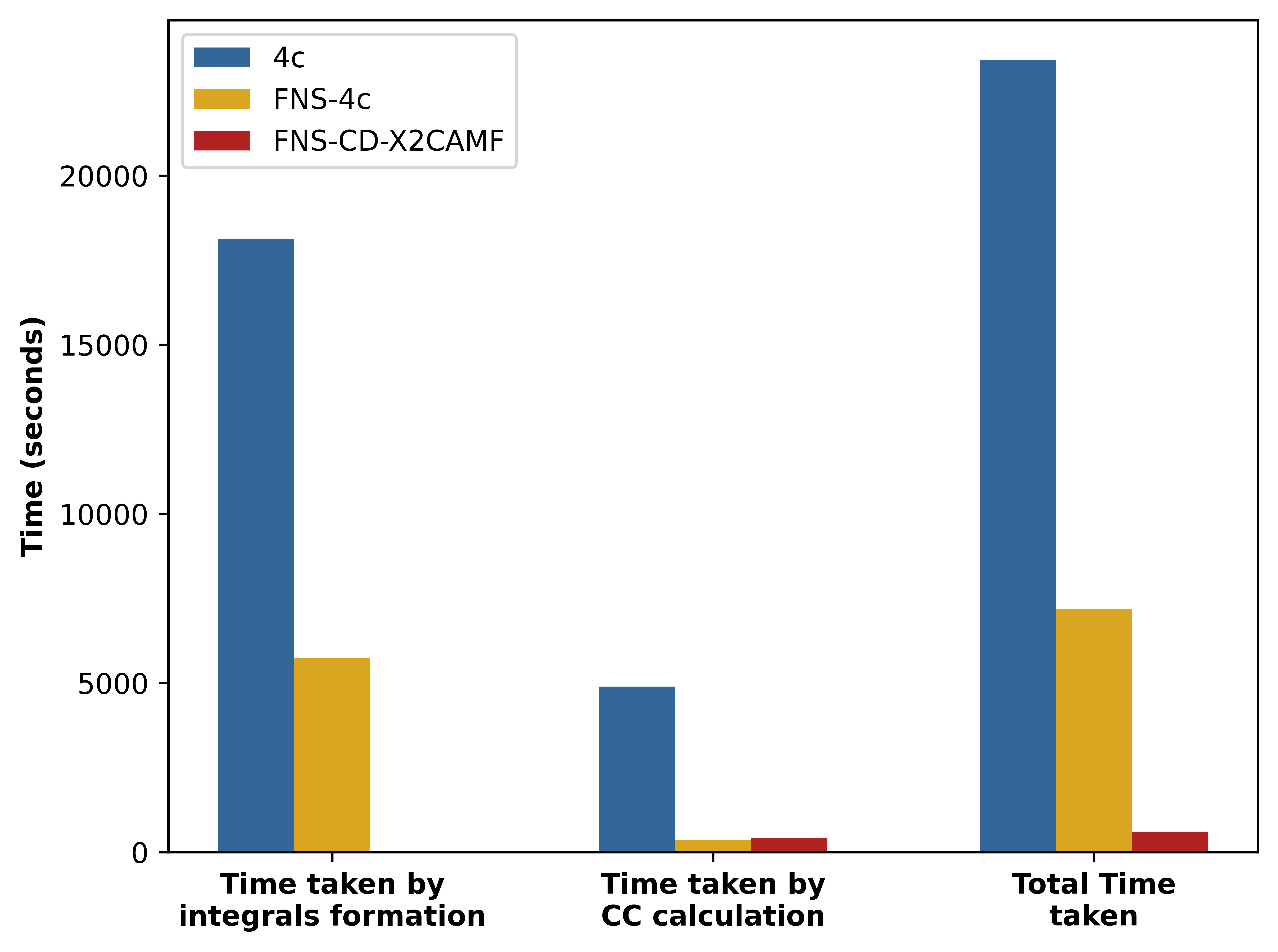}
    \caption{Comparison of the time taken by the different steps in correlation calculation of HI molecule in canonical four-component (4c), FNS four-component (FNS-4c), and FNS-CD-X2CAMF two-component method. The basis set used was dyall.acv3z for the I atom and uncontracted aug-cc-pVTZ for the H atom.}
    \label{fig:my_label4}
    \end{center}
\end{figure}

To compare the computational efficiency between the four-component canonical CCSD (4c), four-component FNS-CCSD (FNS-4c), and the FNS-CD-X2CAMF-CCSD method, we have calculated the CCSD energy for the HI molecule. An uncontracted aug-cc-pVTZ basis set for H atom and dyall.acv3z basis set for the I atom was used for the calculations. The calculations were carried out sequentially on a dedicated workstation with two Intel(R) Xeon(R) Gold 5315Y processors @ 3.20 GHz. The workstation had a total of 512 GB of RAM. The core electrons were kept frozen throughout the correlation calculation. An FNS threshold of 10$^{-5}$ and a tight threshold (FNS threshold: $10^{-5}$, and CD threshold: $10^{-5}$) were used for FNS-4c and FNS-CD-X2CAMF CCSD calculations.
The virtual space consists of  434 virtuals in the canonical spinor basis. However, in the truncated frozen natural spinor basis, this dimension decreases significantly to 174.
Table \ref{table:4} indicates the storage demand of all the MO integrals involving virtual indices and the time taken for their formation in the four-component canonical spinor basis, four-component FNS-basis, and CD-X2CAMF based two-component FNS basis. It is clear from the table \ref{table:4} that a large computational saving in terms of storage can be achieved by using the FNS-CD-X2CAMF scheme, as it avoids the formation/storage of four and three virtual type integrals and leading to the reduction in timing. In the present implementation, the integrals with two-virtual or one-virtual indices have similar storage requirements in FNS-4c and FNS-CD-X2CAMF methods. However, the time required to construct them is much smaller in the FNS-CD-X2CAMF method. 
Figure \ref{fig:my_label4} displays the timings for integral formation, CCSD iteration, and the total time taken in the calculation of the HI molecule. The FNS-based implementation for 4c-CCSD takes just a fraction of the timing for MO integral formation and the CCSD iterations and hence leads to a drastic reduction in the computational cost. The FNS-CD-X2CAMF-CCSD scheme takes almost the same time for the CCSD iterations as the FNS-4c-CCSD method. However, the time required for integral construction in the FNS basis is negligible in the former, which greatly reduces the overall timing. The FNS-CD-X2CAMF-CCSD calculations are $\sim$ 11 times faster than the four-component FNS-CCSD calculation and are $\sim$ 38 times faster than the four-component canonical CCSD calculations, meanwhile lifting the storage bottlenecks as well.

\begin{table}[H]
\centering
\caption{Size of the integral used, and time taken to compute them in canonical four-component (4c), FNS four-component (FNS-4c), and FNS-CD-X2CAMF two-component method for HI molecule. The basis set used was dyall.acv3z for the I atom and uncontracted aug-cc-pVTZ for the H atom.}
\begin{tabular}{ c c c c c c c c }
\hline \hline \\
  & \multicolumn{3}{c}{Storage (GB)} & & \multicolumn{3}{c}{Time taken (seconds)} \\
 \cline{2-4} \cline{6-8} \\
 Integral & 4c    & FNS-4c & FNS-CD-X2CAMF & & 4c    & FNS-4c & FNS-CD-X2CAMF  \\
 \hline \\
 VVVV  & 2150 & 55 & - & & 13407 & 1626 & - \\
 OVVV & 110 & 7 & - & & 1212 & 812 & - \\
 OOVV  & 0.9 & 0.14 & 0.14  & & 659 & 610 & 0.35 \\
 OOOV  & 0.19 & 0.07 & 0.07 & & 865 & 782 & 0.06 \\
\hline \hline 
\end{tabular}
\label{table:4}
\end{table}

\subsection{Application to medium-sized complex}
\begin{figure}[H]
    \begin{center}
    \includegraphics[width=0.5\textwidth]{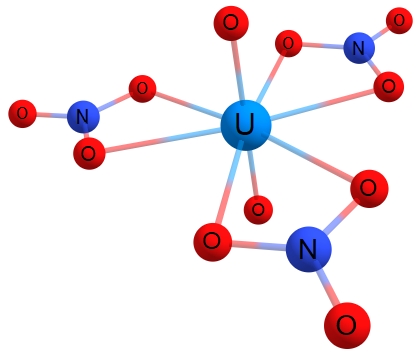}
    \caption{Molecular structure for the [UO$_{2}$(NO$_{3}$)$_{3}$]$^{-}$ complex.}
    \label{fig:my_label5}
    \end{center}
\end{figure}

As a potential application of the FNS-CD-X2CAMF-CCSD method, we performed correlation calculation for [UO$_{2}$(NO$_{3}$)$_{3}$]$^{-}$ complex. The geometry for the complex is taken from the work of Pototschnig and co-workers \cite{doi:10.1021/acs.jctc.1c00260}. The molecular structure of the complex is shown in Figure \ref{fig:my_label5}. An uncontracted
aug-cc-pVDZ basis set for H, O, and N atom and s-aug-dyall.v2z basis set for the U atom was used for the calculations. The [UO$_{2}$(NO$_{3}$)$_{3}$]$^{-}$ complex has a basis set of dimension 1594 with 202 occupied spinors and 1392 virtual spinors. The core electrons were kept frozen through the correlation calculation, and a LOOSEFNS truncation threshold was used for the FNS-CD-X2CAMF-CCSD calculation, resulting in  106 occupied spinors and 562 virtual spinors. The number of Cholesky vectors at the loose threshold is 2441. The calculations were carried out sequentially on a specialized workstation equipped with two Intel(R) Xeon(R) Gold 5315Y processors @ 3.20 GHz and a total of 2.0 TB of RAM. The time taken in Cholesky vector formation in the AO basis is 17 minutes and 17 seconds.
Since the OOVV type integrals in a canonical basis also possess a storage bottleneck for this case, they are also generated on the fly and are directly stored on the memory in the FNS basis.
The time taken by the two-electron integral formations in the FNS basis is 8 minutes and 16 seconds. The CCSD calculation took 2 days, 5 hours, 59 minutes, and 10 seconds. Out of which, 16 hours are taken by the construction of the particle-particle ladder term ( $\sum\limits_{ef} \langle ab \Vert ef \rangle t_{ef}^{ij}$).

\section{Conclusions}
In this study, we propose an efficient FNS-CD-X2CAMF-CCSD/CCSD(T) method that incorporates frozen natural spinors and the Cholesky decomposition technique for two-component X2CAMF-CCSD/CCSD(T) methods.  Our benchmark calculations on gas phase non-covalent ligand dissociation enthalpy of coinage metal ion complexes demonstrate that a normal threshold is sufficient to achieve a balanced trade-off between maintaining chemical accuracy and computational efficiency. The ability of the FNS-CD-X2CAMF-CCSD(T) method to accurately replicate the experimental gas-phase non-covalent ligand dissociation enthalpy at the CBS limit has been investigated, and our calculation shows that a reasonable level of accuracy for calculating bond dissociation enthalpies can be achieved. 
The comparison between the computational efficiency of the CCSD calculations using four-component canonical, four-component FNS, and two-component FNS-CD-X2CAMF methods demonstrates that FNS-CD-X2CAMF-based relativistic calculations are significantly faster than their canonical four-component counterparts as well as outrank them in terms of storage requirement of two-electron integrals. The FNS-CD-X2CAMF framework leads to a significant reduction in the computation cost of the coupled cluster method over the standard 4c-FNS framework both in terms of storage requirements and computational time which can significantly reduce the computational cost. The method's applicability was further supported by a correlation calculation for a medium-sized uranium complex. It shows that the method can be routinely used for accurate relativistic calculations of small molecules, even with modest computational resources. A massively parallel version of the FNS-CD-X2CAMF-CCSD/CCSD(T) method will lead to its widespread adoption, in computational studies related to heavy element-containing systems and complexes.
Work is in progress in that direction.
\begin{acknowledgement}
The authors acknowledge the support from the IIT
Bombay, CRG, and Matrix project of DST-SERB, CSIR-India, DST-Inspire Faculty Fellowship, Prime Minister's
Research Fellowship, ISRO for financial support, IIT
Bombay super computational facility, and C-DAC Super-
computing resources (PARAM Yuva-II, Param Bramha)
for computational time.

\end{acknowledgement}

\begin{suppinfo}
The following file is available free of charge.
\begin{itemize}
  \item SI: The convergence of the FNS-CD-X2CAMF-CCSD and CCSD(T) bond dissociation enthalpies with respect to the Cholesky threshold and the absolute values of the dissociation enthalpies in different bases.
\end{itemize}

\end{suppinfo}

\providecommand{\latin}[1]{#1}
\makeatletter
\providecommand{\doi}
  {\begingroup\let\do\@makeother\dospecials
  \catcode`\{=1 \catcode`\}=2 \doi@aux}
\providecommand{\doi@aux}[1]{\endgroup\texttt{#1}}
\makeatother
\providecommand*\mcitethebibliography{\thebibliography}
\csname @ifundefined\endcsname{endmcitethebibliography}  {\let\endmcitethebibliography\endthebibliography}{}

\end{document}